\documentclass[runningheads]{llncs}
\usepackage{graphicx}
\usepackage{multirow}
\usepackage{amsmath}
\usepackage[misc,geometry]{ifsym} 
\usepackage{booktabs,subcaption,amsfonts,dcolumn}
\newcolumntype{d}[1]{D..{#1}}

\begin{document}
\title{ Learning self-calibrated optic disc and cup segmentation from multi-rater annotations}

\author{Junde Wu\and
Huihui Fang \and
Fangxin Shang \and Zhaowei Wang \and Dalu Yang \and Wenshuo Zhou \and Yehui Yang \and Yanwu Xu\textsuperscript{ (\Letter)}}


\authorrunning{J. Wu et al.}

\institute{Intelligent Healthcare Unit, Baidu Inc.\\
\email{ywxu@ieee.org}
\\}
\maketitle              

\begin{abstract}
The segmentation of optic disc (OD) and optic cup (OC) from fundus images is an important fundamental task for glaucoma diagnosis. In the clinical practice, it is often necessary to collect opinions from multiple experts to obtain the final OD/OC annotation. This clinical routine helps to mitigate the individual bias. But when data is multiply annotated, standard deep learning models will be inapplicable. In this paper, we propose a novel neural network framework to learn OD/OC segmentation from multi-rater annotations. The segmentation results are self-calibrated through the iterative optimization of multi-rater expertness estimation and calibrated OD/OC segmentation. In this way, the proposed method can realize a mutual improvement of both tasks and finally obtain a refined segmentation result. Specifically, we propose Diverging Model (DivM) and Converging Model (ConM) to process the two tasks respectively. ConM segments the raw image based on the multi-rater expertness map provided by DivM. DivM generates multi-rater expertness map from the segmentation mask provided by ConM. The experiment results show that by recurrently running ConM and DivM, the results can be self-calibrated so as to outperform a range of state-of-the-art (SOTA) multi-rater segmentation methods.

\end{abstract}

\section{Introduction}\label{sec:introduction}

Accurate annotation of the optic disc and cup (OD/OC) on fundus image can significantly facilitate the glaucoma diagnosis \cite{garway1998vertical,wu2020leveraging}. With the development of deep learning methods, automated OD/OC segmentation from the fundus images become more and more popular recently \cite{60,21,30,4,6}. Training the deep learning models often requires the a single ground-truth for each instance. However, in order to mitigate the individual bias, it is common to collect the annotations from multiple clinical experts clinically. It makes the deep learning models which work well on nature images can not be directly applied to this task. This problem is called 'multi-rater problem' by the prior works \cite{ji2021learning,yu2020difficulty,wu2022opinions,liao2021modeling}. A common practice toward the problem is to take majority vote, i.e., taking the average of multiple labels. Being simple and easy to implement, this strategy, however, comes at the cost of ignoring the varied expertise-level of multiple experts \cite{fu2020retrospective}.

Recently, the problem of multi-rater labels start to attract research attention. A part of methods are proposed to learn calibrated results which aware the inter-observer variability \cite{guan2018said,chou2019every,jensen2019improving,ji2021learning}. It is shown the calibrated results will achieve better performance on a variety of ground-truths fused from multi-rater labels. However, they still need rater expertness provided to guide the calibration. As a result, in most cases, they are still limited to predict traditional majority vote.

We can see beyond learning the calibrated model which is aware of the uncertainty, estimating the multi-rater expertness indicating which rater is more credible is also necessary. A few previous methods proposed to assign multi-rater expertness based on the prior knowledge that reflects the confidence of the raters \cite{warfield2004simultaneous,raykar2010learning,rodrigues2018deep,albarqouni2016aggnet,tanno2019learning,cao2019max,wu2022opinions}. One general and reliable prior knowledge is the segmentation prior of raw images \cite{raykar2010learning,rodrigues2018deep,albarqouni2016aggnet}. In OD/OC segmentation, it means that labels closer to the OD/OC structure of fundus images should be given more confidence. However, these methods learned fused ground-truth without calibration. Therefore, the multi-rater expertness cannot be dynamically adjusted in the inference stage, which causes the results to be either overconfident or ambiguous.

In order to make up for the shortcomings of these two branches of study, we propose a novel recurrent neural network to jointly calibrate OD/OC segmentation and estimate the multi-rater expertness maps. We dubbed the process as self-calibrated segmentation. In the recurrence, multi-rater expertness maps are provided as a guidance for the calibration and the calibrated masks are again used for the multi-rater expertness evaluation. By the regularization of fundus image prior, the iterative optimization will converge and the two tasks can be mutually improved. Specifically, we propose Converging Model (ConM) and Diverging Model (DivM) to learn the two tasks respectively. ConM learns calibrated OD/OC segmentation based on the multi-rater expertness maps provided by DivM. It is achieved by the feature integration based on the attention mechanism \cite{vaswani2017attention}. DivM learns to separate multi-rater labels from the OD/OC masks provided by ConM. We theoretically prove this separation process is equivalent the estimation of multi-rater expertness. By the iterative optimization of DivM and ConM, it can be proved both calibrated segmentation and multi-rater expertness will converge to the optimal solutions. The experimental results also show the results can be gradually improved with the recurrence of ConM and DivM.

Three major contributions are made with this paper. First, toward the multi-rater OD/OC segmentation, we propose a novel recurrent learning framework for the self-calibrated segmentation. The framework jointly learns calibrated segmentation and estimates multi-rater expertness, which gains mutual improvement on both tasks. Second, in this recurrent learning framework, we propose ConM and DivM for the calibrated segmentation and multi-rater expertness assignment. Attention is adopted to integrate the expertness maps into segmentation decoder. Finally, we validate the proposed method on th OD/OC segmentation. Our method shows superior performance compared with SOTA multi-rater learning strategies.

\section{Theoretical Premises}\label{sec:assumptions}
Suppose that there are $M$ raters, $K$ classes, e.g. {optic disc, optic cup, background} in OD/OC segmentation task. Denote by a matrix $z^{m} \in \mathbb{R}^{H\times W \times K}$ the observed label of rater $m$, $H$ and $W$ are the height and width of the item respectively. Denote by a matrix $w^{m} \in \mathbb{R}^{H\times W \times K}$ the expertness map of rater $m$. Let $z^{[M]}$ denotes ${z^{1},z^{2},...,z^{M}}$. The data point $x$ and multi-rater labels $z^{[M]}$ are assumed to be drawn i.i.d. from random variables $X$ and $Z^{1}, Z^{2}...Z^{M}$. 

Denote by $y$ that the fusion of $z^{[M]}$ by multi-rater expertness maps $w^{[M]}$, which can be expressed as:
\begin{equation}\label{equ:gt}
    y = softmax (\sum_{m = 1}^{M} w^{m} \cdot z^{m} + p), 
\end{equation}
where $\cdot$ represents element-wise multiplication, $p$ is the prior. We softmax the matrix dimension which represents the classes, to make sure the sum of the possibility is 1. Denote by $y^{*}$ the potentially correct label and $w^{*}$ the optimal expertness maps to attain it. There has:

\begin{proposition}\label{pro:1}
If and only if $w^{*m} = log P (z^{m}| y^{*})$, it has $ P (y^{*} | z^{[M]})=$ \\$softmax (\sum_{m = 1}^{M} w^{*m} \cdot z^{m} + p)$.
\end{proposition}
Proposition \ref{pro:1} is proved in supplementary material. It implies learning optimal expertness $w^{*m}$ is equivalent to learn $P (z^{m} | y^{*})$. This enables us to supervise DivM by the observed multi-rater labels $z^{M}$. The estimated masks $\tilde{z}^{[M]}$ of DivM can be used as $w^{m}$ to self-fuse the multi-rater labels:
\begin{equation}\label{equ:gt2}
    y^{self} = \tilde{z}^{[M]} \odot z^{[M]} = softmax (\sum_{m = 1}^{M} log (\tilde{z}^{m}) \cdot z^{m} + p_{u}),
\end{equation}
where $\odot$ denotes the self-fusion operation, $p_{u}$ is the uniform distribution prior, $y^{self}$ is self-fusion label. We construct ConM to estimate $y^{self}$ from raw fundus image and $\tilde{z}^{m}$, and construct DivM to estimate $z^{[M]}$ from $\tilde{y}$ produced by ConM. It can be proved this iterative optimization will converge to the optimal solution $w^{*[M]}$ and $y^{*}$ (shown in supplementary material).

\section{Methodology}

\begin{figure*}[!t]
\centering
\includegraphics[width=1\textwidth]{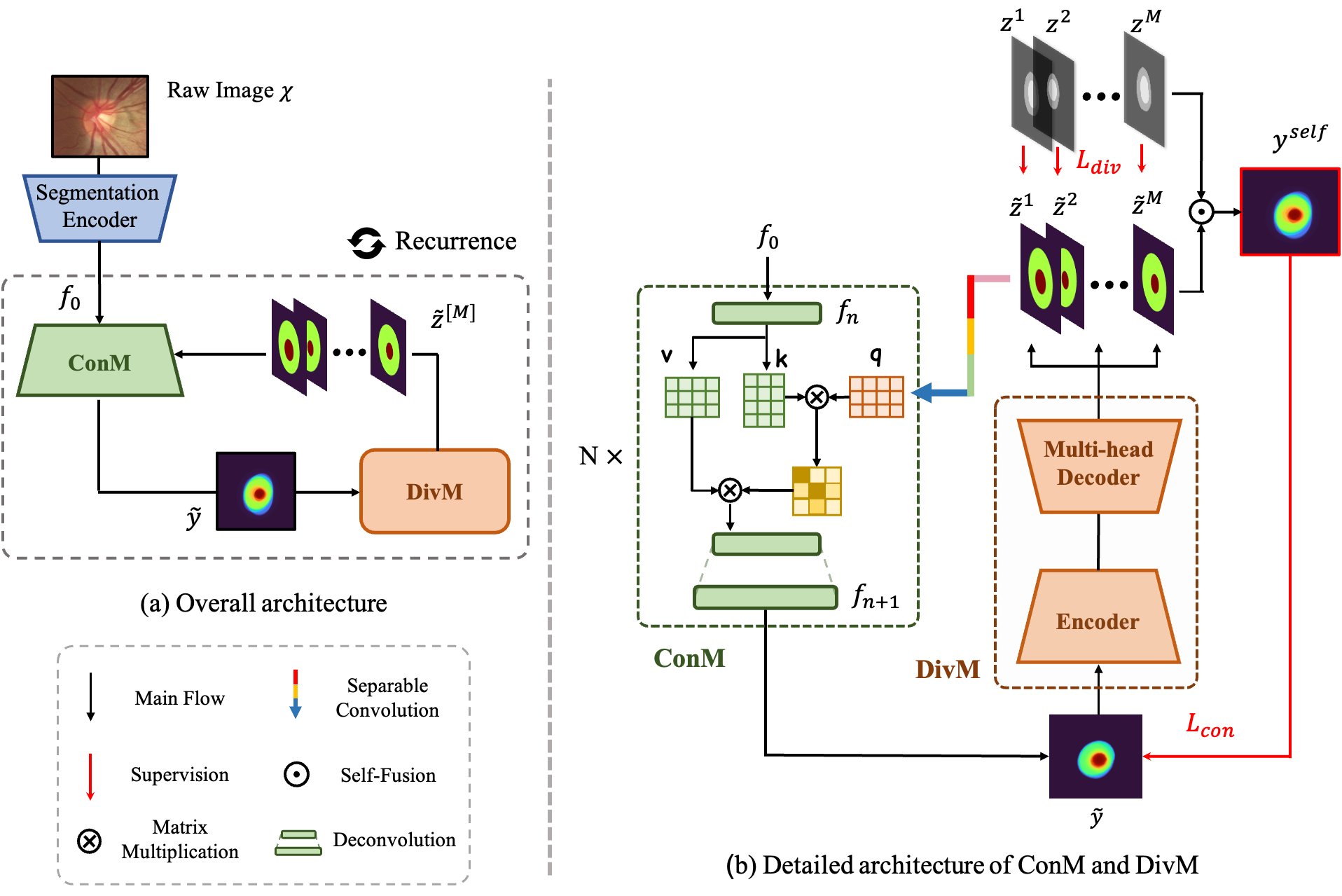}
\caption{Overall architecture of the proposed self-calibration segmentation method. Green denotes ConM modules. Orange denotes DivM modules.}
\label{fig:overall}
\end{figure*}

In this paper, we propose a recurrent model to learn self-calibrated segmentation from multi-rater annotations. The overall flow of the proposed method is shown in Fig. \ref{fig:overall} (a). Raw image $x$ is first sent into a CNN-based encoder to obtain a deep embedding $f_{0} \in \mathbb{R}^{\frac{H}{r} \times \frac{W}{r} \times C}$, where $r$ is the down sample rate, $C$ is the channel number of the embedding. Then ConM will use $f_{0}$ and given expertness maps $\tilde{w}^{[M]}$ to estimate a fused segmentation mask $\tilde{y} = softmax (\sum_{m = 1}^{M} \tilde{w}^{m} \cdot z^{m} + p_{x})$, where $p_{x}$ is the raw image prior implicitly learned by network itself. Obtained $\tilde{y}$ will be taken as potential accurate segmentation mask $y^{*}$ and be sent to DivM. DivM would then separate it to multi-rater segmentation masks, by estimating $P ( z^{[M]} | y^{*})$. Each head of DivM will estimate its corresponding rater's annotation. Multi-rater expertness maps can be represented by these estimated probability maps as shown in Proposition \ref{pro:1}. Multi-rater expertness maps will then be embedded by separable convolution \cite{chollet2017xception} and sent to ConM in the next iteration. ConM and DivM will recurrently run several times until converge.

\subsection{ConM}

We propose ConM to estimate calibrated segmentation mask based on given multi-rater expertness maps. The basic structure is shown in Fig. \ref{fig:overall} (b). The input of ConM is the raw image embedding $f_{0}$ and the output is the segmentation mask $\tilde{y}$. Multi-rater expertness maps $\tilde{w}^{[M]}$ produced by DivM are integrated into ConM by attention mechanism to calibrate the segmentation. In ConM, attention are inserted into each two deconvolution layers. It takes embedded multi-rater expertness as \textit{query}, segmentation features as \textit{key} and \textit{value}. In this way, the segmentation features can be selected and enhanced based on the given multi-rate expertness maps. 

Specifically, consider ConM at the $n^{th}$ layer, the segmentation feature is $f_{n} \in \mathbb{R}^{\frac{H}{r_{n}} \times \frac{W}{r_{n}} \times C_{n}}$. The embedded multi-rater expertness map is $\bar{w}_{n} \in \mathbb{R}^{\frac{H}{r_{n}} \times \frac{W}{r_{n}} \times C_{n}}$. Then $f_{n}$ is transferred by:
\begin{equation}\label{equation:fd}
     \bar{f}_{n} = {\rm Attention} (q, k, v) = {\rm Attention} (\bar{w}_{n} + E_{w},f_{n} + E_{f}, f_{n}).
\end{equation}
where ${\rm Attention} (query, key, value)$ denotes attention mechanism, $E_{w}, E_{f}$ are positional encodings \cite{carion2020end} for expertness embedding and segmentation feature map respectively. Following \cite{dosovitskiy2020image}, we reshape the feature maps into a sequence of flattened patches before the attention. Similarly, $\bar{f}_{n}$ will be reshaped back to $\mathbb{R}^{\frac{H}{r_{n}} \times \frac{W}{r_{n}} \times C_{n}}$ after the attention. Then a deconvolution layer is applied on the transformed feature $\bar{f}_{n}$ to obtain $f_{n+1} \in \mathbb{R}^{\frac{H}{r_{n+1}} \times \frac{W}{r_{n+1}} \times C_{n+1}}$. 
Such a block is multiply stacked to achieve the final output $\tilde{y}$.

\subsection{DivM}

DivM estimates multi-rater OD/OC labels from the segmentation masks provided by ConM. The input of DivM is the estimated segmentation mask $\tilde{y}$. The output of DivM is the estimated multi-rater annotations $\tilde{z}^{[M]}$. DivM is implemented by a standard convolution encoder-decoder network with $M$ heads, where each head estimates one rater's segmentation annotation. 

The estimated probability maps can be represented by the multi-rater expertness maps according Proposition \ref{pro:1}. In order to integrate the expertness into ConM by attention, these maps are embedded to the same size as the target segmentation feature. We use separable convolution \cite{chollet2017xception} which contains a pair of point-wise convolution and depth-wise convolution to embed multi-rater expertness maps. Pointwise convolution keeps the scale of the maps but deepen the channels, while depth-wise convolution downsamples the features but keeps the channel number. These layers not only reshape the maps but also embed the maps to the feature level for the integration.

\subsection{Supervision}

Consider ConM and DivM run once each, i.e, from $f_{0}$ to $\tilde{z}^{[M]}$, as one recurrence. Each instance will run $\tau$ recurrences in a single epoch. We backforward the gradients of the model after $\tau$ times of recurrence. The total loss function is represented as:
\begin{equation}\label{equation:totalloss}
    \mathcal{L}_{total} = \sum_{i = 1}^{\tau} \mathcal{L}_{div}^{i} + \mathcal{L}_{con}^{i},
\end{equation}
where $\mathcal{L}_{con}$ and $\mathcal{L}_{div}$ are the loss functions for ConM and DivM, respectively. 

$\mathcal{L}_{div}$ is the loss function for DivM. The estimation of each head is supervised by the corresponding multi-rater label, which is:
\begin{equation}\label{equation:fd}
    \mathcal{L}_{div} = \sum_{m = 1}^{M} \mathcal{L}_{CE} (\tilde{z}^{m},z^{m}),
\end{equation}
where $\mathcal{L}_{CE}$ denotes the cross-entropy loss function.

$\mathcal{L}_{con}$ constraints ConM to estimate the self-fusion label, which is $y^{self} = softmax (\sum_{m = 1}^{M} log (\tilde{z}^{m}) \cdot z^{m} + p_{u})$. However, note that ConM is supposed to learn the raw image prior $p_{x}$ that cannot be supervised. Therefore, instead of the pixel-level supervision, we adopt Structural Similarity Index (SSIM) as the loss function. 
SSIM constraints the estimated mask to have a similar structure with self-fusion label but also allows the slight difference caused by the raw image prior. Formally, $\mathcal{L}_{con}$ for the $i^{th}$ recurrence is represented as:
\begin{equation}\label{equation:la}
\begin{split}
     \mathcal{L}_{con}^{i} & = \textrm{SSIM} (\tilde{y}_{i}, y^{self}_{i-1}).
\end{split}
\end{equation}

The gradients are backforward individually in each recurrence, which means the gradients of the $i^{th}$ recurrence will not effect the $ (i-1)^{th}$ recurrence.

\section{Experiment}


\subsection{Implement Details}

In the experiments, we initialize the recurrence by the expertness maps sampled from uniform distribution. We use ResUnet \cite{resunet} as the backbone for ConM and DivM. Segmentation encoder is jointly trained with ConM and DivM in the experiments. All the experiments are implemented with the PyTorch platform and trained/tested on 4 Tesla P40 GPU with 24GB of memory. All training and test images are uniformly resized to the dimension of 256$\times$256 pixels. The networks are trained in an end-to-end manner using Adam optimizer \cite{kingma2014adam} with a mini-batch of 16 for 80 epochs. The learning rate is initially set to 1 $\times 10^{-4}$. If not specifically mentioned, we use the results of the $4^{th}$ recurrence for the comparison. The detailed hyper-parameters and model architectures can be found in the supplementary material and open-source code.


\subsection{Main Results}

To verify the proposed model dynamically calibrated the segmentation results in the recurrence, we show the results in different recurrences for the comparison. 
The results are verified on REFUGE \cite{orlando2020refuge} OD/OC segmentation dataset (1200 samples), which is annotated by seven medical experts. The segmentation performance measured by dice score of OD ($\mathcal{D}_{disc}$) and OC ($\mathcal{D}_{cup}$) is shown in Table \ref{tab:allrefuge}. The outputs of DivM are compared with multi-rater labels, and the output of ConM is compared with the self-fusion label. In practice, we conduct four times of recurrence since the results are stable since then. 
\begin{table}[h]

\centering
\setlength\arrayrulewidth{0.1pt}
\caption{The performance of DivM and ConM in different recurrences. DivM is evaluated by multi-rater labels. ConM is evaluated by self-fusion ground-truth.}
\resizebox{\columnwidth}{!}{
\begin{tabular}{c|cccccccccccccc|cc}
\hline
Models & \multicolumn{14}{c|}{DivM}                                                                                                                                                                                                                                                                                                                                                                     & \multicolumn{2}{c}{ConM}         \\ \hline
Raters & \multicolumn{2}{c}{R1}                                 & \multicolumn{2}{c}{R2}                                 & \multicolumn{2}{c}{R3}                                 & \multicolumn{2}{c}{R4}                                 & \multicolumn{2}{c}{R5}                                 & \multicolumn{2}{c}{R6}                                 & \multicolumn{2}{c|}{R7}            & \multicolumn{2}{c}{Self-Fusion}   \\ \hline
Dice   & \multicolumn{1}{c}{$\mathcal{D}_{disc}$}     & \multicolumn{1}{c}{$\mathcal{D}_{cup}$}     & \multicolumn{1}{c}{$\mathcal{D}_{disc}$}     & \multicolumn{1}{c}{$\mathcal{D}_{cup}$}     & \multicolumn{1}{c}{$\mathcal{D}_{disc}$}     & \multicolumn{1}{c}{$\mathcal{D}_{cup}$}     & \multicolumn{1}{c}{$\mathcal{D}_{disc}$}     & \multicolumn{1}{c}{$\mathcal{D}_{cup}$}     & \multicolumn{1}{c}{$\mathcal{D}_{disc}$}     & \multicolumn{1}{c}{$\mathcal{D}_{cup}$}     & \multicolumn{1}{c}{$\mathcal{D}_{disc}$}     & \multicolumn{1}{c}{$\mathcal{D}_{cup}$}     & \multicolumn{1}{c}{$\mathcal{D}_{disc}$}     & $\mathcal{D}_{cup}$     & \multicolumn{1}{c}{$\mathcal{D}_{disc}$}     & $\mathcal{D}_{cup}$     \\ \hline
Rec1   & \multicolumn{1}{c}{96.06} & \multicolumn{1}{c}{84.45} & \multicolumn{1}{c}{95.56} & \multicolumn{1}{c}{83.13} & \multicolumn{1}{c}{94.64} & \multicolumn{1}{c}{81.55} & \multicolumn{1}{c}{\textbf{94.50}} & \multicolumn{1}{c}{77.71} & \multicolumn{1}{c}{95.00} & \multicolumn{1}{c}{87.52} & \multicolumn{1}{c}{\textbf{94.60}} & \multicolumn{1}{c}{\textbf{80.56}} & \multicolumn{1}{c}{\textbf{94.67}} & \textbf{75.45} & \multicolumn{1}{c}{95.48} & 87.80 \\ 
Rec2   & \multicolumn{1}{c}{96.14} & \multicolumn{1}{c}{85.05} & \multicolumn{1}{c}{95.35} & \multicolumn{1}{c}{85.28} & \multicolumn{1}{c}{\textbf{94.71}} & \multicolumn{1}{c}{81.00} & \multicolumn{1}{c}{94.33} & \multicolumn{1}{c}{80.46} & \multicolumn{1}{c}{95.62} & \multicolumn{1}{c}{\textbf{88.07}} & \multicolumn{1}{c}{94.11} & \multicolumn{1}{c}{78.12} & \multicolumn{1}{c}{94.28} & 74.44 & \multicolumn{1}{c}{95.60} & 88.63 \\
Rec3   & \multicolumn{1}{c}{96.58} & \multicolumn{1}{c}{86.27} & \multicolumn{1}{c}{96.20} & \multicolumn{1}{c}{86.17} & \multicolumn{1}{c}{94.42} & \multicolumn{1}{c}{\textbf{81.63}} & \multicolumn{1}{c}{94.35} & \multicolumn{1}{c}{81.12} & \multicolumn{1}{c}{95.42} & \multicolumn{1}{c}{86.76} & \multicolumn{1}{c}{94.30} & \multicolumn{1}{c}{77.39} & \multicolumn{1}{c}{94.66} & 73.48 & \multicolumn{1}{c}{96.24} & 89.55 \\
Rec4   & \multicolumn{1}{c}{\textbf{96.77}} & \multicolumn{1}{c}{\textbf{86.49}} & \multicolumn{1}{c}{\textbf{96.39}} & \multicolumn{1}{c}{\textbf{86.39}} & \multicolumn{1}{c}{94.53} & \multicolumn{1}{c}{81.19} & \multicolumn{1}{c}{94.41} & \multicolumn{1}{c}{\textbf{81.34}} & \multicolumn{1}{c}{\textbf{95.89}} & \multicolumn{1}{c}{86.72} & \multicolumn{1}{c}{94.14} & \multicolumn{1}{c}{76.88} & \multicolumn{1}{c}{94.46} & 73.30 & \multicolumn{1}{c}{\textbf{96.33}} & \textbf{89.71} \\ 
\bottomrule
\end{tabular}%
}
\label{tab:allrefuge}
\end{table}

In Table \ref{tab:allrefuge}, we can see the performance of DivM keeps increasing on Rater1, Rater2 and Rater4 (R1, R2, R4) in the recurrence, while dropping on Rater5, Rater6 and Rater7 (R5, R6, R7). It indicates the model consistently calibrates the results in the recurrence. Constraint by the raw image prior, some raters become more and more credible in the recurrence. We can also see the results of ConM are gradually improved on self-fusion labels, indicating the increasing consistency of DivM and ConM with the recurrence. Both the results of DivM and ConM become stable in the $4^{th}$ recurrence, indicating the results have converged.

\begin{figure}[h]
\centering
\includegraphics[width=0.95\textwidth]{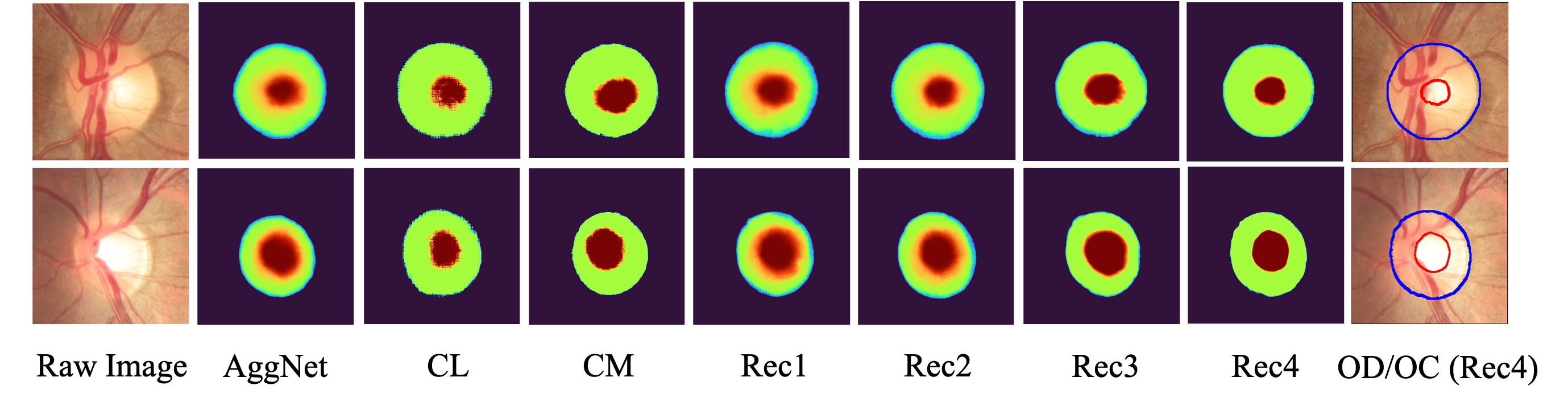}
\caption{Visualized comparison of OD/OC segmentation. Column 2-4 are the results of AggNet \cite{albarqouni2016aggnet}, CL \cite{rodrigues2018deep} and CM \cite{tanno2019learning}, respectively. Column 5-9 are self-calibrated results in recurrence 1, 2, 3, 4 and the final OD/OC boundary, respectively.}
\label{fig:vis}
\end{figure}

We also show the visualization results in Fig. \ref{fig:vis}. We find the model predicts uncertain masks first, then calibrate the results to be more confident through the recurrence, to an end of confident and calibrated result.

\begin{table}[h]
\centering
\setlength\arrayrulewidth{0.1pt}
\caption{Quantitative comparison results of SOTA calibrated/non-calibrated segmentation methods. Results are measured by Dice score (\%).}
\begin{subtable}[t]{0.8\textwidth}
\centering
\resizebox{\columnwidth}{!}{
\begin{tabular}{c|c|cc|cc|cc|cc|cc}
\toprule
     &  & \multicolumn{2}{c|}{MV} & \multicolumn{2}{c|}{STAPLE} & \multicolumn{2}{c|}{LFC} & \multicolumn{2}{c|}{Diag} & \multicolumn{2}{c}{Self Fusion} \\ \hline
      &     & $\mathcal{D}_{disc}$     & $\mathcal{D}_{cup}$      & $\mathcal{D}_{disc}$       & $\mathcal{D}_{cup}$          & $\mathcal{D}_{disc}$      & $\mathcal{D}_{cup}$          & $\mathcal{D}_{disc}$       & $\mathcal{D}_{cup}$           & $\mathcal{D}_{disc}$      & $\mathcal{D}_{cup}$         \\ \hline
\multirow{3}{*}{No Calibrated} & AGNet    & 90.21  & 71.86  & 89.45   & 70.30     & 88.40  & 69.65          & 88.98  & 68.57   & 91.26 & 72.83 \\
 & pOSAL    & 94.52  & 83.81  & 93.97   & 83.08     & 93.06  & 81.86         & 92.82  & 80.32    & 94.30 & 83.05 \\ 
 &BEAL     & 94.84  & 84.92  & 94.28   & 84.14     & 93.49  & 83.27            & 92.88  & 81.93    & 95.14 & 84.66 \\ \hline
 \multirow{3}{*}{Calibrated} & WDNet    & 94.63  & 84.46  & 95.32   & 84.32     & 93.14  & 80.55          & 92.53  & 81.03   & 94.67 & 83.60 \\
 & UECNN    & 94.42  & 84.80  & 94.53   & 84.26     & 94.22  & 82.79          & 94.07  & 80.71   & 95.47 & 84.87 \\
 & MRNet    & 95.00  & 86.40  & 94.13   & 85.25     & 93.71  & 84.14 & 93.53   & 82.14    & 95.43 & 86.42 \\ \cline{1-12}
 Self-calibrate & Ours & \textbf{95.23}  & \textbf{87.75}  & \textbf{95.34}   & \textbf{88.14}     & \textbf{94.67}  & \textbf{86.82}            & \textbf{94.20}  & \textbf{85.25}   & \textbf{96.33} & \textbf{89.71}\\ \hline
\bottomrule
\end{tabular}%
}
\caption{\footnotesize Performance on REFUGE dataset.}
\label{tab:sotarefuge}
\end{subtable}
\hspace{\fill}
\begin{subtable}[t]{0.8\textwidth}
\centering
\resizebox{\columnwidth}{!}{
\begin{tabular}{c|c|cc|cc|cc|cc|cc}
\toprule
       &    & \multicolumn{2}{c|}{MV} & \multicolumn{2}{c|}{STAPLE} & \multicolumn{2}{c|}{LFC} & \multicolumn{2}{c|}{Diag} & \multicolumn{2}{c}{Self Fusion} \\ \hline
     &      & $\mathcal{D}_{disc}$     & $\mathcal{D}_{cup}$      & $\mathcal{D}_{disc}$       & $\mathcal{D}_{cup}$          & $\mathcal{D}_{disc}$      & $\mathcal{D}_{cup}$          & $\mathcal{D}_{disc}$       & $\mathcal{D}_{cup}$           & $\mathcal{D}_{disc}$      & $\mathcal{D}_{cup}$         \\ \hline
\multirow{3}{*}{No Calibrated}&  AGNet    & 96.31  & 78.05  & 95.30   & 77.02         & 95.37   & 75.82     & 95.23  & 72.45   & 95.32 & 77.40 \\
 & pOSAL    & 95.85  & 84.07  & 95.37   & 83.25        & 95.07   & 83.75     & 96.22  & 79.28    & 96.10 & 85.61 \\
 & BEAL     & 97.08  & 85.97  & 96.13   & 84.16         & 95.87   & 84.82     & 95.06  & 81.06    & 96.28 & 86.89 \\ \hline
\multirow{3}{*}{Calibrated} & WDNet    & 96.81  & 82.17  & 96.15   & 81.25         & 96.11   & 82.78     & 95.18  & 80.31   & 95.53 & 84.38 \\
 & UECNN    & 96.37  & 85.52  & 95.88   & 83.59    & 96.16  & 84.34   & 95.60  & 82.73          & 95.48  & 85.92    \\
 & MRNet    & \textbf{97.55}  & 87.20  & 96.26   & 86.37         & \textbf{96.58}
 & 85.77    & 95.11  & 82.55    & 96.33 & 86.83 \\ \cline{1-12}
 Self-calibrate & Ours & 97.22  & \textbf{88.24}  & \textbf{96.34}   & \textbf{88.50}    & 96.23   & \textbf{88.48}     & \textbf{96.63}  & \textbf{85.78}   & \textbf{97.82} & \textbf{90.15} \\ \hline
\bottomrule
\end{tabular}
}
\caption{\footnotesize Performance on RIGA dataset.}
\label{tab:sotariga}
\end{subtable}
\end{table}

\begin{table}[h]
\caption{Ablation study on attentive integration and SSIM loss function}
\centering
\resizebox{0.8\columnwidth}{!}{
\begin{tabular}{cc|cccccccc}
\hline
    \multirow{2}{*}{SSIM} & \multirow{2}{*}{\begin{tabular}[c]{@{}c@{}}Attentive\\ Integration\end{tabular}}& \multicolumn{2}{c}{MV}                                            & \multicolumn{2}{c}{STAPLE}                                        & \multicolumn{2}{c}{LFC}                                           & \multicolumn{2}{c}{Diag}                                                                      \\ \cline{3-10} 
  &  & $\mathcal{D}_{disc}$            & $\mathcal{D}_{cup}$             & $\mathcal{D}_{disc}$            & $\mathcal{D}_{cup}$             & $\mathcal{D}_{disc}$            & $\mathcal{D}_{cup}$             & $\mathcal{D}_{disc}$            & $\mathcal{D}_{cup}$                       \\ \hline
          &      & 94.51                           & 84.12                           & 94.58                           & 84.31                           & 94.33                           & 84.03                           & 93.26                           & 81.48                                                 \\
       \checkmark   &      & 94.72                           & 84.73                           & 95.03                           & 84.90                           & 94.56                           & 84.61                           & 93.77                           & 81.70                                                  \\
          &  \checkmark    & 94.89                           & 85.71                           & 95.35                           & 85.41                           & 94.38                           & 85.27                           & 93.96                           & 82.63                                                   \\ \hline
     \checkmark     &   \checkmark   & \textbf{95.23} & \textbf{87.75} & \textbf{95.34} & \textbf{88.14} & \textbf{94.67} & \textbf{86.82} & \textbf{94.20} & \textbf{85.25}  \\ \hline
\end{tabular}}
\label{tab:integrate}
\end{table}
\subsection{Compared with SOTA}
To verify the calibration ability of the proposed method, we compare it with SOTA calibrated/non-calibrated segmentation methods on REFUGE and RIGA \cite{almazroa2017agreement} dataset. RIGA is a OD/OC segmentation dataset contains 750 samples annotated by six raters. The non-calibrated methods we compare include AGNet \cite{acnet}, BEAL \cite{beal} and  pOSAL \cite{posal}. The calibrated methods we compare include WDNet \cite{guan2018said}, UECNN \cite{jensen2019improving} and MRNet \cite{ji2021learning}. The models are measured on different combinations of multi-rater labels, including majority vote (MV), STAPLE \cite{warfield2004simultaneous}, LFC \cite{raykar2010learning}, and Diag \cite{wu2022opinions} to verify the generalization. The proposed model is supervised by the target label in the first recurrence for the initialization. The quantitative results are shown in Table \ref{tab:sotarefuge} and \ref{tab:sotariga}. We can see calibrated methods basically work better than non-calibrated one. Self-calibrated segmentation consistently achieves superior performance on various multi-rater ground-truths, indicating the better generalization of the proposed method. The performance improvement is especially prominent for OC segmentation where the inter-observer variability is more significant, with an increase of 1.83\% and 1.72\% Dice coefficient over the second on REFUGE-MV and RIGA-MV, respectively. In addition, we also note most segmentation models can achieve better performance on the self-fusion ground-truth, which demonstrates the self-fusion label is easier to learn on account of the awareness of raw image prior.

Not only the multi-rater expertness helps to improve calibration, the calibration can also facilitate the multi-rater expertness estimation. We compare the self-calibration with SOTA self-fusion supervised segmentation methods, including CL \cite{rodrigues2018deep}, CM \cite{tanno2019learning} and AggNet \cite{albarqouni2016aggnet}. The visual comparisons is shown in Fig. \ref{fig:vis}. We can see the CL and CM who explicitly learn the multi-rater expertness, will be overconfident to the inaccurate results, while AggNet who implicitly learns the multi-rater expertness is prone to obtain ambiguous results. The proposed self-calibrated segmentation is able to estimate the result from uncertain to confident with the recurrence, to an end of confident and calibrated result.

\subsection{Ablation study}
Ablation studies are performed over attentive integration and SSIM loss function on REFUGE dataset, as listed in Table \ref{tab:integrate}. The experiments are evaluated on a range of ground-truths. Simple feature concatenation is used to replace attention in the ablation study. In Table \ref{tab:integrate}, as we sequentially adding SSIM and attention, the model performance is gradually improved. It also shows the combination usage of SSIM and attention boosts the performance more than the individual components, indicating the awareness of fundus image prior and multi-rater expertness are mutually improved.

\section{Conclusion}
Toward learning OD/OC segmentation from multi-rater labels, we propose a self-calibrated segmentation model to recurrently calibrated the segmentation and estimate the multi-rater expertness. In this way, the shortcomings of the two independent tasks are complemented, thus gain the mutual improvement. Extensive empirical experiments demonstrated the self-calibrated segmentation outperforms both the calibrated segmentation methods and expertness-aware segmentation methods. Future works will continue to exploit the potential of the proposed method and extend it to the other multi-rater segmentation tasks.
\bibliographystyle{splncs04}
\bibliography{egbib}
\clearpage

\end{document}